\documentclass{ws-ijmpa}

\def\x{\chi}

\def\e{\epsilon}

\def\m{m_{\rm th}}

\def\be{\begin{equation}}
\def\ee{\end{equation}}

\def\lsim{\raise0.3ex\hbox{$<$\kern-0.75em\raise-1.1ex\hbox{$\sim$}}}
\def\gsim{\raise0.3ex\hbox{$>$\kern-0.75em\raise-1.1ex\hbox{$\sim$}}}

\def\NP{{ Nucl.\ Phys.\ }}
\def\PL{{ Phys.\ Lett.\ }}
\def\PR{{ Phys.\ Rev.\ }}

\def\ZP{{ Z.\ Phys.\ }}
\def\EP{{ Europ.\ Phys.\ J.\ }}

\begin{document}

\markboth{Helmut Satz}
{Critical Behaviour in Statistical QCD}

%
\catchline{}{}{}{}{}
%

\title{CRITICAL BEHAVIOUR IN STATISTICAL QCD}

\author{\footnotesize HELMUT SATZ\footnote{also at
CFIF, Instituto Superior T\'ecnico, Lisbon, Portugal.}}

\address{Department of Physics, Bielefeld University, Universit{\"a}tsstrasse
25\\
D-33615 Bielefeld, Germany
}



\maketitle

\pub{Received (Day Month Year)}{Revised (Day Month Year)}

\begin{abstract}

We survey the critical behaviour of strongly interacting matter, as
determined in finite temperature lattice QCD.

\end{abstract}

\section{Phases of Strongly Interacting Matter} 

What happens to strongly interacting matter in the limit of
high temperature and/or density? This question has fascinated
physicists ever since short range strong interactions and the
resulting multiple hadron production were discovered. Let us
look at some of the features that have emerged.

\begin{figure}[htb]
\centerline{\psfig{file=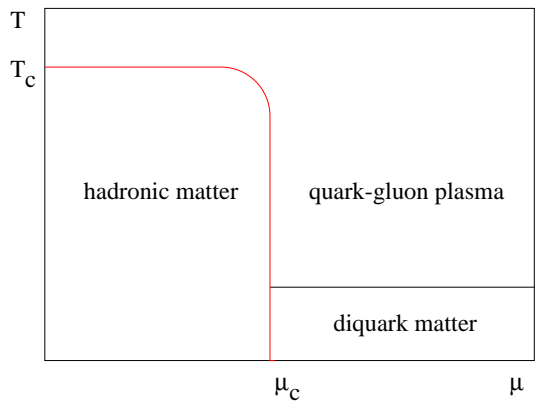,width=7cm}}
\caption{The phase diagram of QCD} 
\label{phase}
\end{figure}

\begin{itemize}
\item{Hadrons have an intrinsic size, with a radius $r_h \simeq
1$ fm, and hence a hadron needs a space of volume $V_h \simeq (4\pi
/3)r_h^3$ in order to exist. This suggests a limiting
density $n_c$ of hadronic matter\cite{Pomeranchuk}, with $n_c = 1/V_h
\simeq 1.5~n_0$, where $n_0 \simeq 0.17$ fm$^{-3}$ denotes the 
density of normal nuclear matter.}
\item{Hadronic interactions provide abundant resonance production, and 
the resulting number $\rho(m)$ of hadron species increases
exponentially as function of the resonance mass $m$,
$\rho(m) \sim \exp(b~\!m)$. Such a form for $\rho(m)$ appeared first
in the statistical bootstrap model, based on self-similar resonance
formation or decay\cite{Hagedorn}. It was then also obtained in the 
more dynamical dual resonance approach\cite{DRM}. In hadron
thermodynamics, the exponential increase of the resonance degeneracy results 
in an upper limit for the temperature of hadronic matter\cite{Hagedorn},
$T_c = 1/b \simeq 150 - 200$ MeV.}
\item{What happens beyond $T_c$? In QCD, hadrons are dimensionful 
color-neutral bound states of more basic pointlike colored quarks and
gluons. Hadronic matter, consisting of colorless constituents of hadronic
dimensions, can therefore turn at high temperatures and/or densities
into a quark-gluon plasma of pointlike colored quarks and gluons as 
constituents\cite{C-P}. This deconfinement transition leads to a 
colour-conducting
state and thus is the QCD counterpart of the insulator-conductor transition
in atomic matter\cite{HS-Fort}.}
\item{A shift in the effective constituent mass is a second transition
phenomenon expected from the behavior of atomic matter. At $T=0$, 
in vacuum, quarks dress themselves with gluons to form the constituent 
quarks that make up hadrons. As a result, the bare quark mass $m_q \sim 0$
is replaced by a constituent quark mass $M_q \sim 300$ MeV. In a
hot medium, this dressing melts and $M_q \to 0$. Since the QCD
Lagrangian for $m_q=0$ is chirally symmetric, $M_q \not= 0$ implies
spontaneous chiral symmetry breaking. The melting $M_q \to 0$ thus
corresponds to chiral symmetry restoration. We shall see later on
that in QCD, as in atomic physics, the shift of the constituent
mass coincides with the onset of conductivity.} 
\item{A third type of transition would set in if the attractive
interaction between quarks leads in the deconfined phase to the 
formation of colored bosonic diquark pairs, the Cooper pairs of QCD.
These diquarks can then condense at low temperature to form a color
superconductor. Heating will dissociate the diquark pairs and turn
the color superconductor into a normal color conductor.}
\end{itemize}
Using the baryochemical potential $\mu$ as a measure for the
baryon density of the system, we thus expect the phase diagram 
of QCD to have the schematic form shown in Fig.\ \ref{phase}. 
Given QCD as the fundamental theory of strong interactions, we
can use the QCD Lagrangian as dynamics input to derive the resulting 
thermodynamics of strongly interacting matter. For vanishing 
baryochemical potential, $\mu=0$, this can be evaluated with the
help of the lattice regularisation, leading to finite temperature
lattice QCD. 

\section{Finite Temperature Lattice QCD}

QCD is defined by the Lagrangian
\be
{\cal L}~=~-{1\over 4}F^a_{\mu\nu}F^{\mu\nu}_a~-~\sum_f{\bar\psi}^f
_\alpha(i \gamma^{\mu}\partial_{\mu} + m_f
-g \gamma^{\mu}A_{\mu})^{\alpha\beta}\psi^f_\beta
~,\label{2.4}
\ee
with
\be
F^a_{\mu\nu}~=~(\partial_{\mu}A^a_{\nu}-\partial_{\nu}A^a_{\mu}-
gf^a_{bc}A^b_{\mu}A^c_{\nu})~. \label{2.5}
\ee
Here $A^a_{\mu}$ denotes the gluon field of colour $a$ ($a$=1,2,...,8)
and $\psi^f_{\alpha}$ the quark field of colour $\alpha$
($\alpha$=1,2,3) and flavour $f$; the input (`bare') quark masses are
given by $m_f$. With the dynamics thus determined, the corresponding
thermodynamics is obtained from the partition function, which is
most suitably expressed as a functional path integral,
\be
Z(T,V) = \int ~dA~d\psi~d{\bar\psi}~
\exp~\left(-\int_V d^3x \int_0^{1/T} d\tau~
{\cal L}(A,\psi,{\bar\psi})~\right), \label{2.6}
\ee
since this form involves directly the Lagrangian density defining the
theory. The spatial integration in the exponent of Eq.\ (\ref{2.6}) is
performed over the entire spatial volume $V$ of the system; in the
thermodynamic limit it becomes infinite. The time component $x_0$ is
``rotated" to become purely imaginary, $\tau = ix_0$, thus turning the
Minkowski manifold, on which the fields $A$ and $\psi$ are originally
defined, into a Euclidean space. The integration over $\tau$ in Eq.\
(\ref{2.6}) runs over a finite slice whose thickness is determined by
the temperature of the system.

From $Z(T,V)$, all thermodynamical observables can be calculated in
the usual fashion. Thus
\be
\epsilon = (T^2/V)\left({\partial \ln Z \over \partial T}\right)_V
\label{2.7}
\ee
gives the energy density, and
\be
P = T \left({\partial \ln Z\over \partial V}\right)_T
\label{2.8}
\ee
the pressure. For the study of critical behaviour, long range
correlations and multi-particle interactions are of crucial importance;
hence perturbation theory cannot be used. The necessary
non-perturbative regularisation scheme is provided by the lattice
formulation of QCD\cite{Wilson}; it leads to a form which can be
evaluated numerically by computer simulation\cite{Creutz}.

The first variable to consider is the
deconfinement measure given by the Polyakov loop\cite{Larry,Kuti}
\be
L(T) \sim \lim_{r \to \infty}~\exp\{-V(r)/T\} \label{2.9}
\ee
where $V(r)$ is the potential between a static quark-antiquark pair
separated by a distance $r$. In the limit of large input quark mass,
$V(\infty)= \infty$ in the confined phase, so that then $L=0$.
Colour screening, on the other hand, makes $V(r)$ finite at large $r$,
so that in the deconfined phase, $L$ does not vanish. It thus becomes
an `order parameter' like the magnetisation in the Ising model.
In the large quark mass limit, QCD reduces to pure $SU(3)$ gauge theory, 
which is invariant under a global $Z_3$ symmetry. The Polyakov loop provides 
a measure of the state of the system under this symmetry: it vanishes for 
$Z_3$ symmetric states and becomes finite when $Z_3$ is spontaneously 
broken. Hence the critical behaviour of $SU(3)$ gauge theory is in the 
same universality class as that of $Z_3$ spin theory (the 3-state Potts 
model): both are due to the spontaneous symmetry breaking of a global 
$Z_3$ symmetry\cite{Svetitsky}.

For finite quark mass $m_q$, $V(r)$ remains finite for $r \to \infty$,
since the `string' between the two colour charges `breaks' when the
corresponding potential energy becomes equal to the mass $M_h$ of the
lowest hadron; beyond this point, it becomes energetically more
favourable to produce an additional hadron. Hence now $L$ no longer
vanishes in the confined phase, but only becomes exponentially small
there,
\be
L(T) \sim \exp\{-M_h/T\}; \label{2.10}
\ee
here $M_h$ is of the order of the $\rho$-mass, so that $L \sim
10^{-2}$, rather than zero. Deconfinement is thus indeed much like the
insulator-conductor transition, for which the order parameter, the
conductivity $\sigma(T)$, also does not really vanish for $T>0$, but
with $\sigma(T) \sim \exp\{-\Delta E/T\}$ is only exponentially small,
since thermal ionisation (with ionisation energy $\Delta E$) produces
a small number of unbound electrons even in the insulator phase.

\begin{figure}[htb]
\vspace*{-2cm}
\mbox{
\epsfig{file=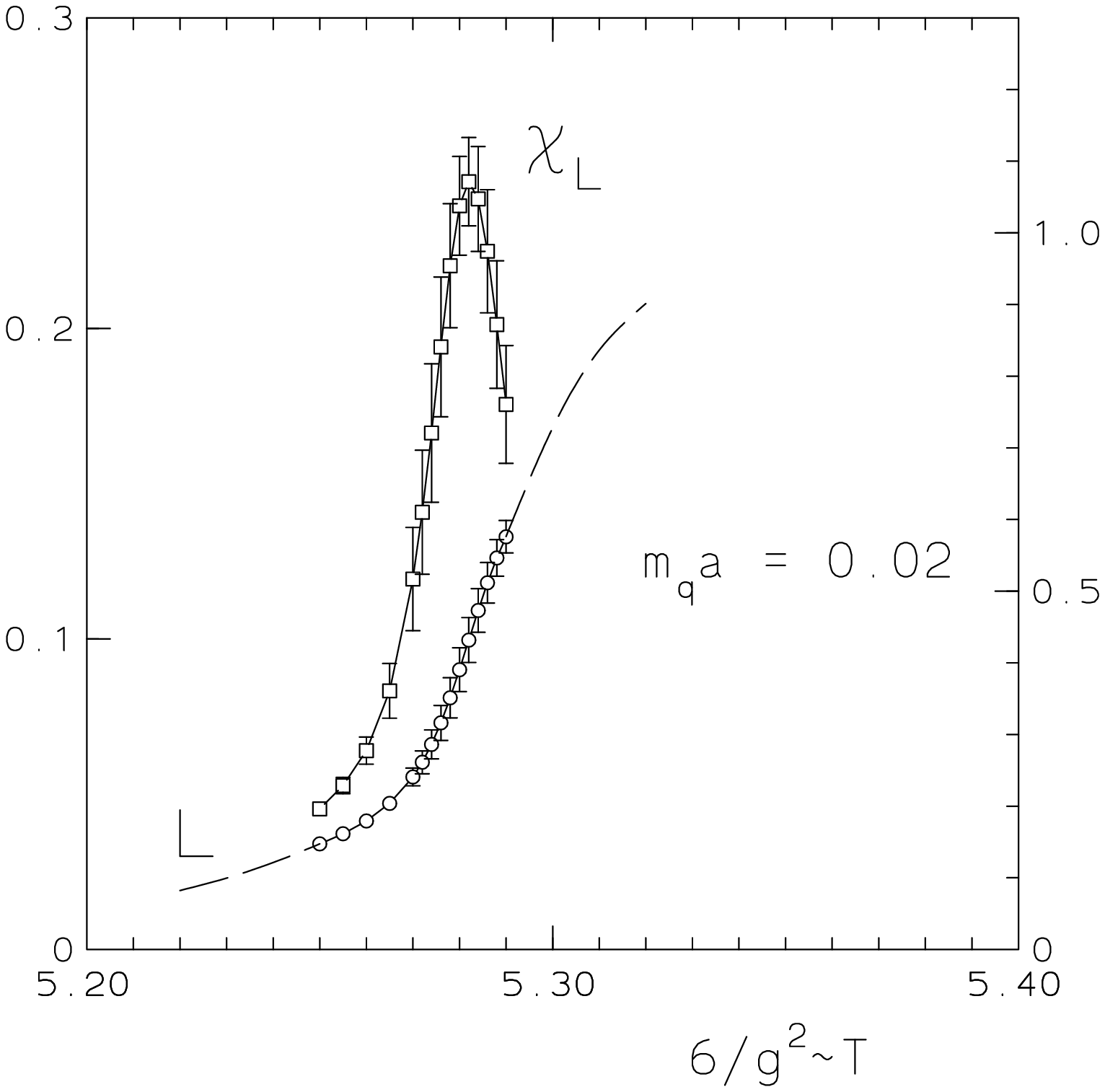,width=6cm}
\hspace*{0.5cm}
\epsfig{file=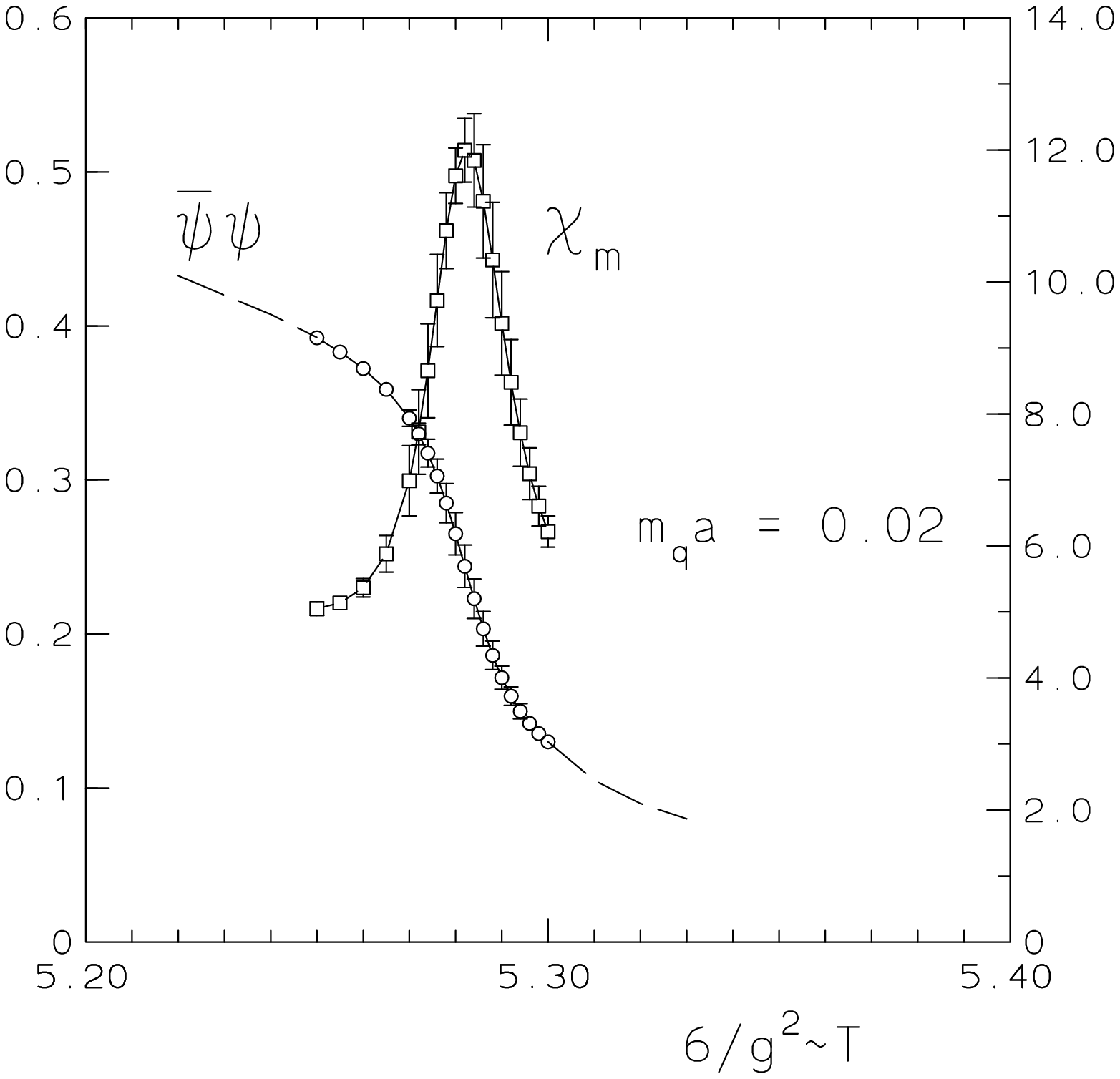,width=6cm}}
\vskip-1.6cm
\hspace*{2.7cm} (a) \hskip 6.5cm (b) 
\caption{Polyakov loop and chiral condensate in two-flavour 
QCD\protect\cite{K&L}} 
\label{2_4}
\end{figure}

Fig.\ \ref{2_4}$~\!$a shows recent lattice results\cite{K&L}
 for $L(T)$ and the
corresponding susceptibility $\x_L(T) \sim \langle L^2 \rangle - 
\langle L \rangle^2$. The calculations were performed 
for the case of two flavours of light quarks, using a current quark 
mass about four times larger than that needed for the physical pion mass.
We note that $L(T)$ undergoes the expected sudden increase from a small
confinement to a much larger deconfinement value. The sharp peak of
$\chi_L(T)$ defines quite well the transition temperature $T_c$, which
we shall shortly specify in physical units.

The next quantity to consider is the effective quark mass; it is
measured by the expectation value of the corresponding term in the
Lagrangian, $\langle {\bar \psi} \psi \rangle(T)$. In the
limit of vanishing current quark mass, the Lagrangian becomes chirally
symmetric and $\langle {\bar \psi} \psi \rangle(T)$ the corresponding
order parameter. In the confined phase, with effective constituent quark
masses $M_q \simeq 0.3$ GeV, this chiral symmetry is
spontaneously broken, while in the deconfined phase, at high enough
temperature, we expect its restoration. In the real world, with finite
pion and hence finite current quark mass, this symmetry is also only
approximate, since $\langle {\bar \psi} \psi \rangle (T)$ now never
vanishes at finite $T$.

The behaviour of $\langle {\bar \psi} \psi \rangle(T)$ and the
corresponding susceptibility $\chi_m \sim \partial \langle {\bar \psi}
\psi \rangle / \partial m_q$ are shown in Fig.\ \ref{2_4}$~\!$b\cite{K&L},
calculated for the same case as above in Fig.\ \ref{2_4}$~\!$a. 
We note here the
expected sudden drop of the effective quark mass and the associated
sharp peak in the susceptibility. The temperature at which this occurs
coincides with that obtained through the deconfinement measure. We
therefore conclude that at vanishing baryon number density, quark
deconfinement and the shift from constituent to current quark mass
coincide. 
We thus obtain for $\mu_B=0$ a rather well defined phase structure,
consisting of a confined phase for $T < T_c$, with $L(T) \simeq 0$ and
$\langle {\bar \psi} \psi \rangle(T) \not= 0$, and a
deconfined phase for $T>T_c$ with $L(T)\not= 0$ and
$\langle {\bar \psi} \psi \rangle(T) \simeq 0$. The
underlying symmetries associated to the critical behaviour at $T=T_c$,
the $Z_3$ symmetry of deconfinement and the chiral symmetry of the quark
mass shift, become exact in the limits $m_q \to \infty$ and $m_q \to
0$, respectively. In the real world, both symmetries are only
approximate; nevertheless, we see from Fig.\ \ref{2_4} 
that both associated measures retain an almost critical behaviour.

\begin{figure}[h]
\vskip0.5cm
\begin{minipage}[t]{5.5cm}
\vspace*{-1cm}
\hskip0.1cm 
\epsfig{file=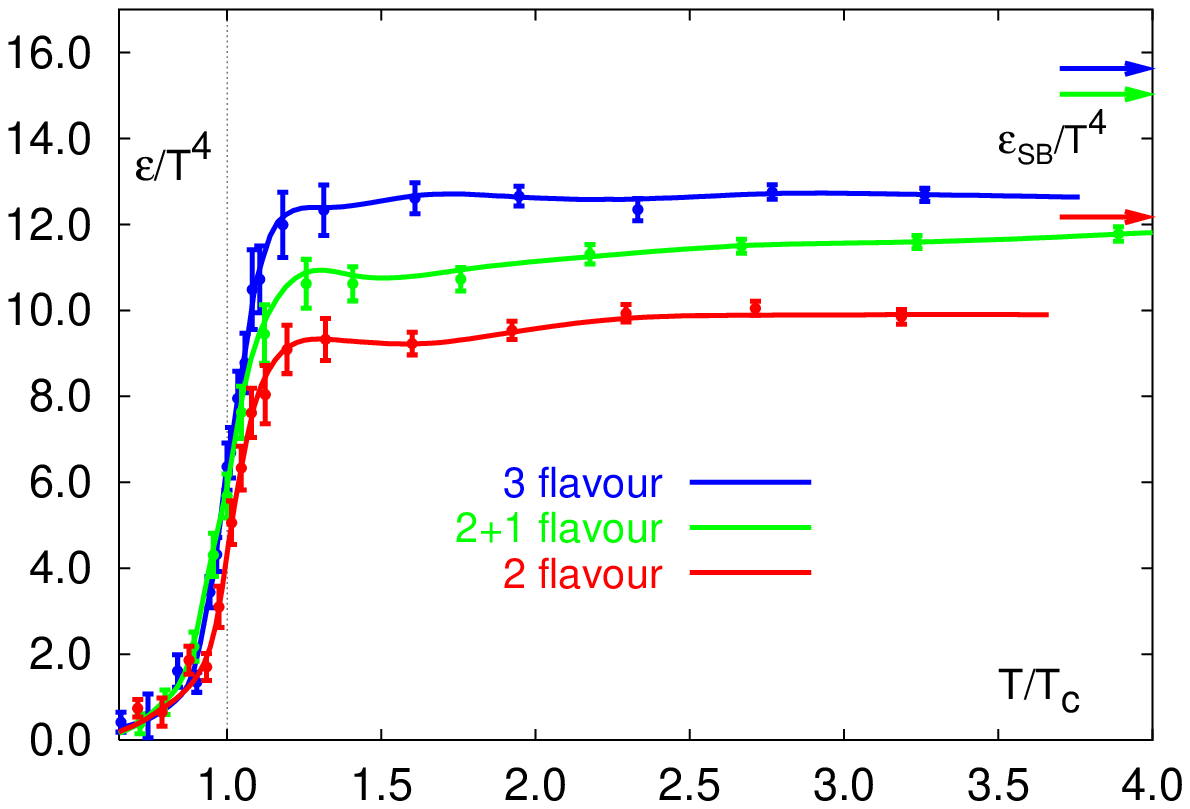,width=5.5cm}
\caption{Energy density vs. temperature\protect\cite{Biele}}
\label{edens}
\end{minipage}
\hspace{1cm}
\begin{minipage}[t]{5.5cm}
\vspace*{-1cm}
\hspace*{-0.2cm}
\epsfig{file=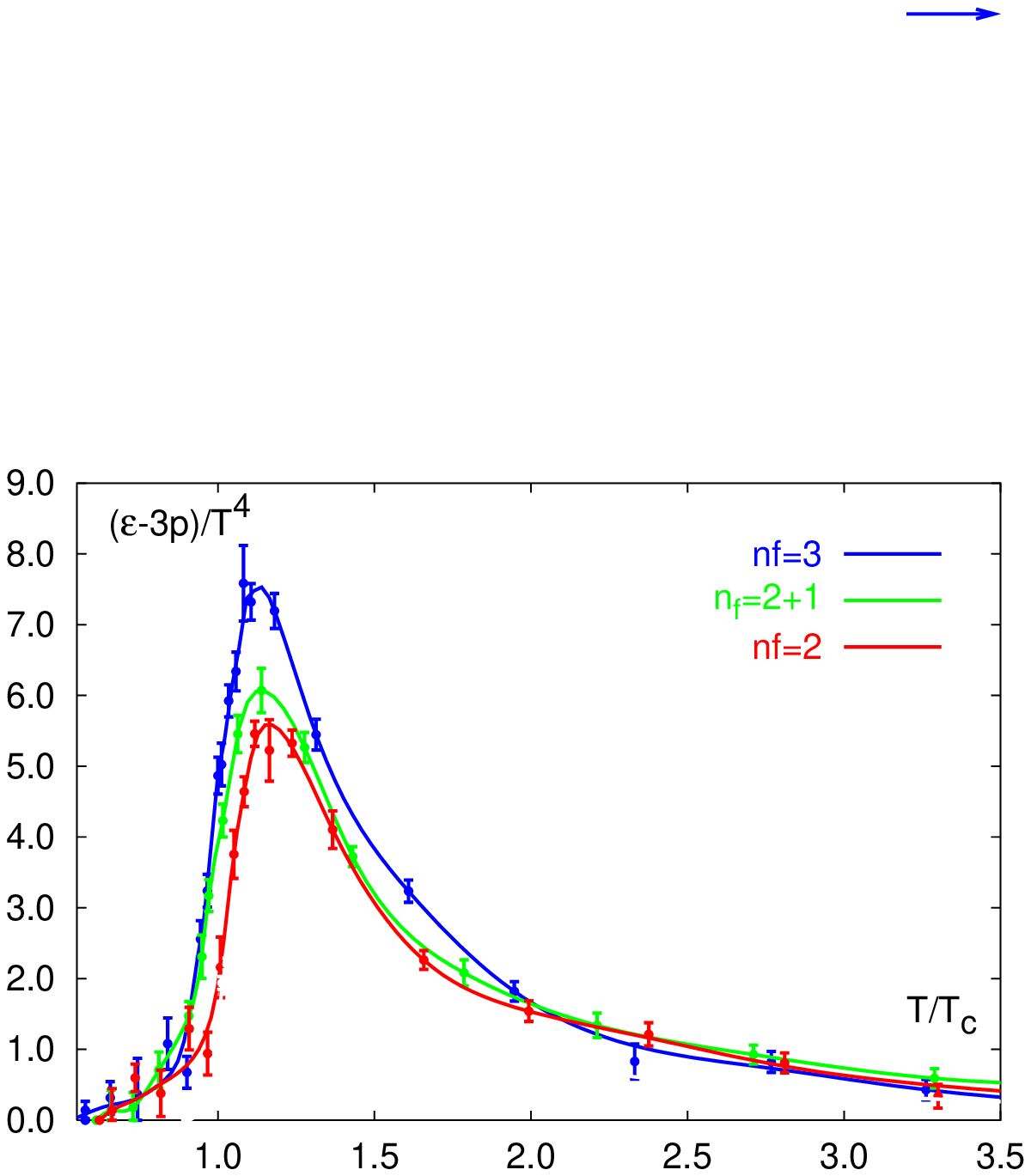,width=5.5cm}
\caption{Interaction measure vs. temperature\protect\cite{delta}} 
\label{inter}
\end{minipage}
\end{figure}

Next we come to the behaviour of energy density $\e$ and pressure $P$ at
deconfinement. In Fig.\ \ref{edens}, it is seen that $\e/T^4$ changes quite
abruptly at the above determined critical temperature $T=T_c$,
increasing from a low hadronic value to one slightly below that
expected for an ideal gas of massless quarks and gluons\cite{Biele}. 


Besides the sudden increase at deconfinement, there are two further points 
to note. In the region $T_c\!<\! T\! <\! 2~T_c$, there still remain strong
interaction effects. As seen in Fig.\ \ref{inter}, the `interaction measure' 
$\Delta=(\e - 3P)/T^4$ remains sizeable and does not vanish, as it would 
for an ideal gas of massless constituents\cite{delta}.
In the simple model of the 
previous section, such an effect arose due to the bag pressure, and in actual 
QCD, one can also interpret it in such a fashion\cite{asakawa}. It has
also been considered in terms of a gradual onset of deconfinement starting 
from high momenta \cite{interaction}, and most recently as a possible 
consequence of coloured ``resonance'' states\cite{Shuryak}.
The second point to note is that the thermodynamic observables remain 
about {10\%} below their Stefan-Boltzmann values (marked ``SB'' in Fig.\ 
\ref{edens}) even at very high temperatures, where the interaction measure
becomes very small. Such deviations from ideal gas 
behaviour can be expressed to a large extent in terms of effective `thermal' 
masses $m_{\rm th}$ of quarks and gluons, with $\m \sim g(T)~T$
\cite{ERSW,Golo,Patkos}. Maintaining the next-to-leading order term
in mass in the Stefan-Boltzmann form gives for the pressure
\be
P = c~\!T^4 \left[1 - a\left({\m\over T}\right)^2\right] = 
c~T^4 [1 - a~g^2(T)]
\ee
and for the energy density 
\begin{eqnarray}
\e & = & 3~\!c~\!T^4 \left[1 - {a\over 3}\left({\m\over T}\right)^2 -
{2a\over 3}\left({\m\over T}\right)\left({d\m\over dT}\right)\right]\nonumber\\
& = & 3~\!c~\!T^4 \left[1 - a~\!g^2(T)
+ {2 a~\! \m \over 3} \left({dg\over dT}\right)\right],
\end{eqnarray}
where $c$ and $a$ are colour- and flavour-dependent positive constants.
The deviations of $P$ and $\e$ from the massless Stefan-Boltzmann form thus 
vanish only logarithmically, with $g(T)\sim 1/\ln~T$, while the 
interaction measure 
\be
\Delta 
= 2~\!a~\!c~\! g(T)~ T {dg\over dT}
\ee
vanishes up to terms of order $T dg/dT$.

Finally we turn to the value of the transition temperature. Since QCD
(in the limit of massless quarks) does not contain any dimensional
parameters, $T_c$ can only be obtained in physical units by expressing
it in terms of some other known observable which can also be calculated
on the lattice, such as the $\rho$-mass, the proton mass, or the string
tension. In the continuum limit, all different ways should lead to the
same result. Within the present accuracy, they define the uncertainty
so far still inherent in the lattice evaluation of QCD. Using the
$\rho$-mass to fix the scale leads to $T_c\simeq 0.15$ GeV, while
the string tension still allows values as large as $T_c \simeq 0.20$
GeV. This means that energy densities of some 1 - 2 GeV/fm$^3$ are
needed in order to produce a medium of deconfined quarks and gluons.

In summary, finite temperature lattice QCD at vanishing baryon density shows
\begin{itemize}
\item{that there is a transition leading to colour deconfinement
coincident with chiral symmetry restoration at $T_c \simeq$ 0.15 - 0.20 GeV;}
\item{that this transition is accompanied by a sudden increase in
the energy density  (the ``latent heat of deconfinement") from a small
hadronic value to a much larger value, about 10\% below
that of an ideal quark-gluon plasma.}
\end{itemize}
In the following section, we want to address in more detail the
nature of the critical behaviour encountered at the transition.

\section{The Nature of the Transition}

We begin with the behaviour for vanishing baryon density ($\mu=0$) and 
come to $\mu\not=0$ at the end. Consider the case of three quark species, 
$u,~d,~s$.
\begin{itemize}
\item{In the limit $m_q \to \infty$ for all quark species, we recover 
pure $SU(3)$ gauge theory, with a deconfinement phase transition provided 
by spontaneous $Z_3$ breaking. It is first order, as is the case for the 
corresponding spin system, the 3-state Potts model.}
\item{For $m_q \to 0$ for all quark masses, the Lagrangian becomes 
chirally symmetric, so that we have a phase transition 
corresponding to chiral symmetry restoration. In the case of three
massless quarks, the transition is also of first order.}

\begin{figure}[htb]
\centerline{\psfig{file=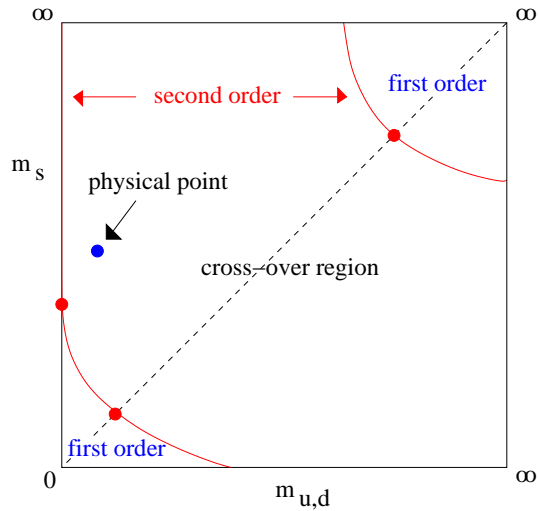,width=7cm}}
\caption{The nature of thermal critical behaviour in QCD}
\label{nature}
\end{figure}

\item{For $0 < m_q < \infty$, there is neither spontaneous $Z_3$
breaking nor chiral symmetry restoration. Hence in general, there is
no singular behaviour, apart from the transient disappearence of the
first order discontinuities on a line of second order transitions.
Beyond this, there is no genuine phase transition, but
only a ``rapid cross-over'' from confinement
to deconfinement. The overall behaviour is summarized in Fig.\ \ref{nature}.} 
\item{As already implicitely noted above, both ``order parameters''
$L(T)$ and $\chi(T)$ nevertheless
show a sharp temperature variation for all values
of $m_q$, so that it is in fact possible to define quite well a 
common cross-over point $T_c$.}
\item{The nature of the transition thus depends quite sensitively
on the number of flavours $N_f$ and the quark mass values: it can
be a genuine phase transition (first order or continuous), or just 
a rapid cross-over. The case realized in nature, the ``physical point'', 
corresponds to small $u,~d$ masses and a larger $s$-quark mass. 
It is fairly certain today that this point falls into the cross-over region.}
\item{Finally we consider briefly the case of finite baryon density,
$\mu\not=0$ \cite{Stephanov}.
Here the conventional computer algorithms of lattice QCD break down,
and hence new calculation methods have to be developed. First such attempts
(reweighting\cite{Fodor}, analytic continuation\cite{Lombardo}, 
power series\cite{Swansea}) suggest the phase diagram shown
in Fig.\ \ref{baryon}. The position of the critical end-point of
the line of first order transition depends on the position of the
physical point in the $m_{u,d}-m_s$ plane, as illustrated in 
Fig.\ \ref{nature}.}

\begin{figure}[htb]
\centerline{\psfig{file=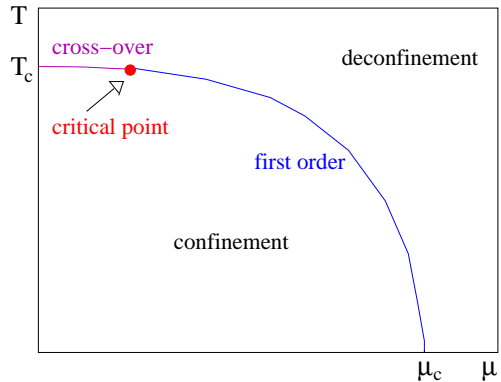,width=6.5cm}}
\caption{Deconfinement in the $T,\mu$ plane}
\label{baryon}
\end{figure}
\vspace*{-0.2cm}
\end{itemize}
The critical behaviour for strongly interacting matter at low or
vanishing baryon density, describing the onset of confinement in 
the early universe and in high energy nuclear collisions, thus occurs 
in the rather enigmatic form of a ``rapid cross-over''. There is no 
thermal singularity and hence, in a strict sense, there are neither 
distinct states of matter nor phase transitions between them. So what 
does the often mentioned experimental search for a ``new state of matter'' 
really mean? How can a new state appear without a phase transition?
Is there a more general way to define and distinguish different states of bulk
media? After all, in statistical QCD one does find that thermodynamic 
observables -- energy and entropy densities, pressure, as well as 
the ``order parameters'' $L(T)$ and $\chi(T)$ -- continue to change rapidly 
and thus define a rather clear transition line in the entire cross-over 
region. Why is this so, what is the mechanism which causes 
such a transition?

\begin{figure}[htb]
\vspace*{0.2cm}
\centerline{\epsfig{file=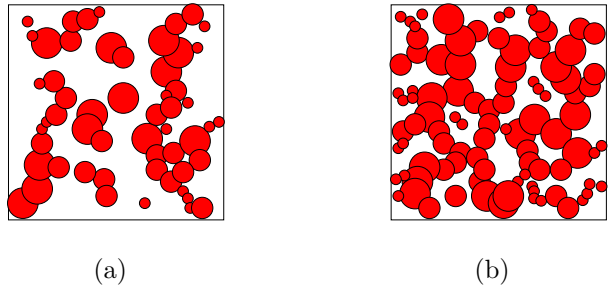,width=8cm}}
\vskip0.3cm
\hskip 3.5cm (a) \hskip 4.5cm (b)
\caption{Hadron and vacuum percolation thresholds}
\label{percotrans}
\end{figure}

In closing this section, we consider a speculative answer to this
rather fundamental question. It is known that in spin systems,
when the partition function becomes analytic and hence there is no more 
thermal critical behaviour, the onset of large cluster formation,
i.e., percolation, persists as a geometric form of critical behaviour 
persits\cite{HS-perco}. This can be applied also to the question of
deconfinement\cite{Magas}. 
Consider hadrons of intrinsic
size $V_h=(4\pi/3)r_h^3$, with $r_h \simeq 0.8$ fm. In three-dimensional 
space, the formation of a connected large-scale cluster first occurs at 
the density
\be
n_c= {0.34 \over V_h} \simeq 0.16~{\rm fm}^{-3}.
\label{hadronmatter} 
\ee
This point specifies the onset of hadronic {\sl matter}, in contrast to
a {\sl gas} of hadrons, and it indeed correctly reproduces the density of 
normal nuclear matter. However, at this density the vacuum as connected 
medium also still exists (see Fig.\ \ref{percotrans}a). 

To prevent infinite connecting vacuum clusters, a much higher hadron 
density is needed. Measured in hadronic size units, the vacuum disappears 
for
\be
\bar{n}_c= {1.24 \over V_h} \simeq 0.56~{\rm fm}^{-3}, 
\label{vacuummatter} 
\ee
schematically illustrated in Fig.\ \ref{percotrans}b. If we assume that at
this point, the medium is of an ideal gas of all known hadrons and hadronic 
resonances, then we can calculate the temperature of the gas at the 
density $\bar{n}_c$: $n_{\rm res}(T_c) = \bar{n}_c$ implies $T_c \simeq 
170$ MeV, which agrees with the value of the deconfinement temperature
found in lattice QCD for $\mu=0$.

We can thus use percolation to define the states of hadronic matter.
At low density, we have a hadron gas, which at the percolation point
$n_c$ turns into connected hadronic matter. When this becomes so dense
that only isolated vacuum bubbles survive, at $\bar{n}_c$, it turns into 
a quark-gluon plasma. This approach provides the correct values both for 
the density of standard nuclear matter and for the deconfinement transition 
temperature.

Such considerations may in fact well be of a more general nature than
the problem of states and transitions in strong interaction physics. 
The question of whether symmetry or connectivity (cluster formation)
determines the different states of many-body systems has intrigued
theorists in statistical physics for quite a long time\cite{Fisher}. 
The lesson learned from spin systems appears to be that cluster formation
and the associated critical behaviour are the more general feature, which 
under certain conditions can also lead to thermal criticality, i.e., 
singular behaviour of the partition function.

\end{document}